\documentclass[11pt]{article}
\usepackage{moriond2000,epsfig}
\def\be{\begin{equation}}
\def\ee{\end{equation}}
\def\bea{\begin{eqnarray}}
\def\eea{\end{eqnarray}}

\begin{document}
\vspace*{4cm}
\title{NUMERICAL STUDY OF THE COSMIC SHEAR}

\author{T. HAMANA$^1$, S. COLOMBI$^{1,2}$, and Y. MELLIER$^{1,3}$}

\address{${}^1$ Institut d'Astrophysique de Paris, CNRS, 
98bis Boulevard Arago, 75014 PARIS, FRANCE\\
${}^2$ {\cal NIC} (Numerical Investigations in Cosmology) Group, CNRS\\
${}^3$ Observatoire de Paris, DEMIRM, 61 avenue de l'Observatoire,
75014 PARIS, FRANCE}

\maketitle

\abstracts{
We study cosmic shear statistics using the ray-tracing simulation
combined with a set of large $N$-body simulations.
In this contribution, we first describe our method.
Then we show some selected results especially focusing on effects of the 
deflection of light rays and the lens-lens coupling which are
neglected in making the theoretical predictions of the cosmic shear
statistics.}

\section{Introduction}
The cosmic shear statistics have been known as a powerful tool for
probing the large-scale structure formation as well as for placing
a constraint on values of the cosmological model \cite{ME99}.
Recently, four independent groups have reported detections of cosmic
shear variance \cite{WA00,BA00,WI00,KA00}.
Although those detections were made with relatively small fields which
limit their accuracy, on going wide field cosmic shear
surveys will provide a precious measurement of not only the cosmic shear
variance but also higher order statistics such like the skewness of 
the lensing convergence.

Since the pioneering work by Gunn \cite{GU67}, there has been a great
progress in the theoretical study of the cosmic shear
statistics \cite{BS00}. 
The analytical formulae for making the theoretical prediction of the
cosmic shear statistics are based on the perturbation
theory of the cosmic density field combined with the nonlinear
clustering ansatz.
The accuracy and limitations of the theoretical predictions should be
tested against numerical simulations that is one
purpose of numerical studies \cite{JA00,WH00}. 

In this contribution, we summarize, briefly, the methods and selected
results of our research project on the numerical study of the cosmic
shear, details are presented in \cite{HA00a,WA00b}.
The project aims (1) to test the analytical predictions against the
numerical simulation, (2) to examine the higher order statistics of the
cosmic shear, (3) to simulate a cosmic shear observation to examine
possible systematic effects caused by, e.g., the source clustering
\cite{BE98,HA00c}.

\section{Models and methods}
\begin{table}
\begin{center}
\caption{Parameters in Three Cluster normalized CDM Models}
\label{tab:cdm}
\begin{tabular}{ccccc}
\hline
\hline
{} & $\Omega_m$  & $\Omega_\lambda$ & $\sigma_8$ & $h$\\
\hline
SCDM & 1.0 & 0.0 & 0.6 & 0.5 \\ 
OCDM & 0.3 & 0.0 & 0.85 & 0.7 \\ 
$\Lambda$CDM & 0.3 & 0.7 & 0.9 & 0.7 \\ 
\hline
\end{tabular}
\end{center}
\end{table}

\begin{figure}
\centerline{
\epsfig{figure=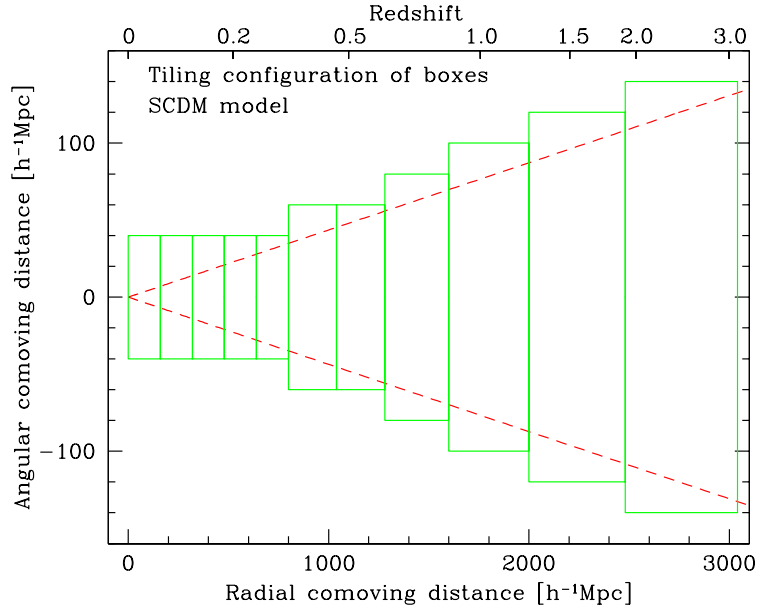,height=8cm}
}
\caption{Tiling configuration of simulation boxes. Dashed lines
shows the comoving angular diameter distance of $\pm$2.5 degrees. 
\label{fig:tiling}}
\end{figure}

We consider three cluster normalized cold dark matter (CDM) models,
parameters in the models are summarized in Table \ref{tab:cdm}.
$N$-body simulations were performed with a vectorized particle-mesh code.
They use $256^2\times512$ particles and the same number of force mesh in a
periodic rectangular comoving box and use light-cone output
\cite{HA00a,HA00b}.
In order to generate the density field from $z=0$ to $z\sim3$, 
we performed 11, 12 and 13 independent simulations for SCDM, OCDM and
$\Lambda$CDM model, respectively.
We adopted the {\it tiling} configuration of the boxes \cite{WH00},
i.e., the box size of each realization is chosen so that we have a
field of view of $5\times5$ square degrees (see Figure \ref{fig:tiling}).

Light ray trajectories are followed through the density field
generated by $N$-body simulations adopting the
multiple-lens plane algorithm \cite{SC92}. 
The lens planes are located at intervals of $80h^{-1}$. 
The initial ray directions are set on $512^2$ grids with the grid spacing of 
$5^\circ/512\sim 0.59$ arcmin.
For each ray, positions of the ray on each lens plane are computed, 
and then the lensing magnification matrix, $M_{ij}$, is computed at
the ray position on each plane.
The lensing convergence, shear and net rotation are expressed by
$\kappa=(M_{11}+M_{22})/2$, $\gamma_1=(M_{11}-M_{22})/2 $,
$\gamma_2=(M_{12}+M_{21})/2$, and $\omega=(M_{12}-M_{21})/2$,
respectively.
We performed 40 realizations for each model changing the underlying
density field (i.e., making random shifts of boxes to $x$ and $y$
directions (perpendicular to the line-of-sight) using the periodic
boundary condition in $N$-body simulations). 

\begin{figure}
\begin{center}
\epsfig{figure=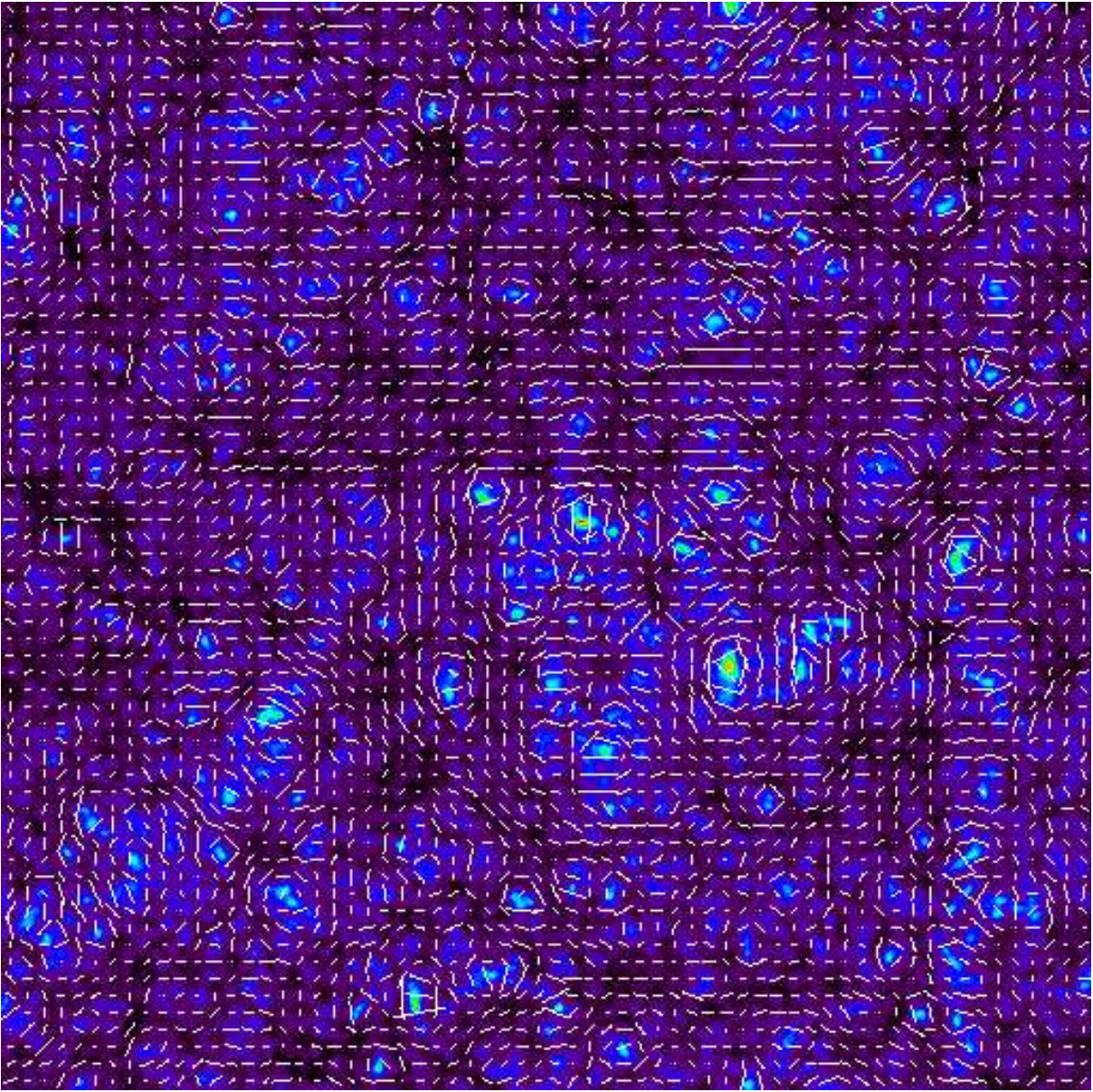,height=16cm}
\caption{The lensing shear map overlaid on the
convergence map for $\Lambda$CDM model, 
the sources are assumed to be at a single redshift of $z_s=1$.
The filed is 5-degree on a side and contains $512^2$ line-of-sights.}
\label{fig:map}
\end{center}
\end{figure}

Figure \ref{fig:map} shows the lensing shear map overlaid on the
convergence map for $\Lambda$CDM model,
the sources are assumed to be at a single redshift of $z_s=1$.
The filed is 5-degree on a side and contains $512^2$ line-of-sights.
The angular resolution is limited by the spatial resolution of
$N$-body simulation.
We found that the effective resolution is about 2 arcmin for the
source redshift of $z_s=1$ and is slightly better (worse) for the
higher (lower) redshift. 

\section{Results and discussion}
\begin{figure}
\begin{minipage}{7.5cm}
\epsfig{figure=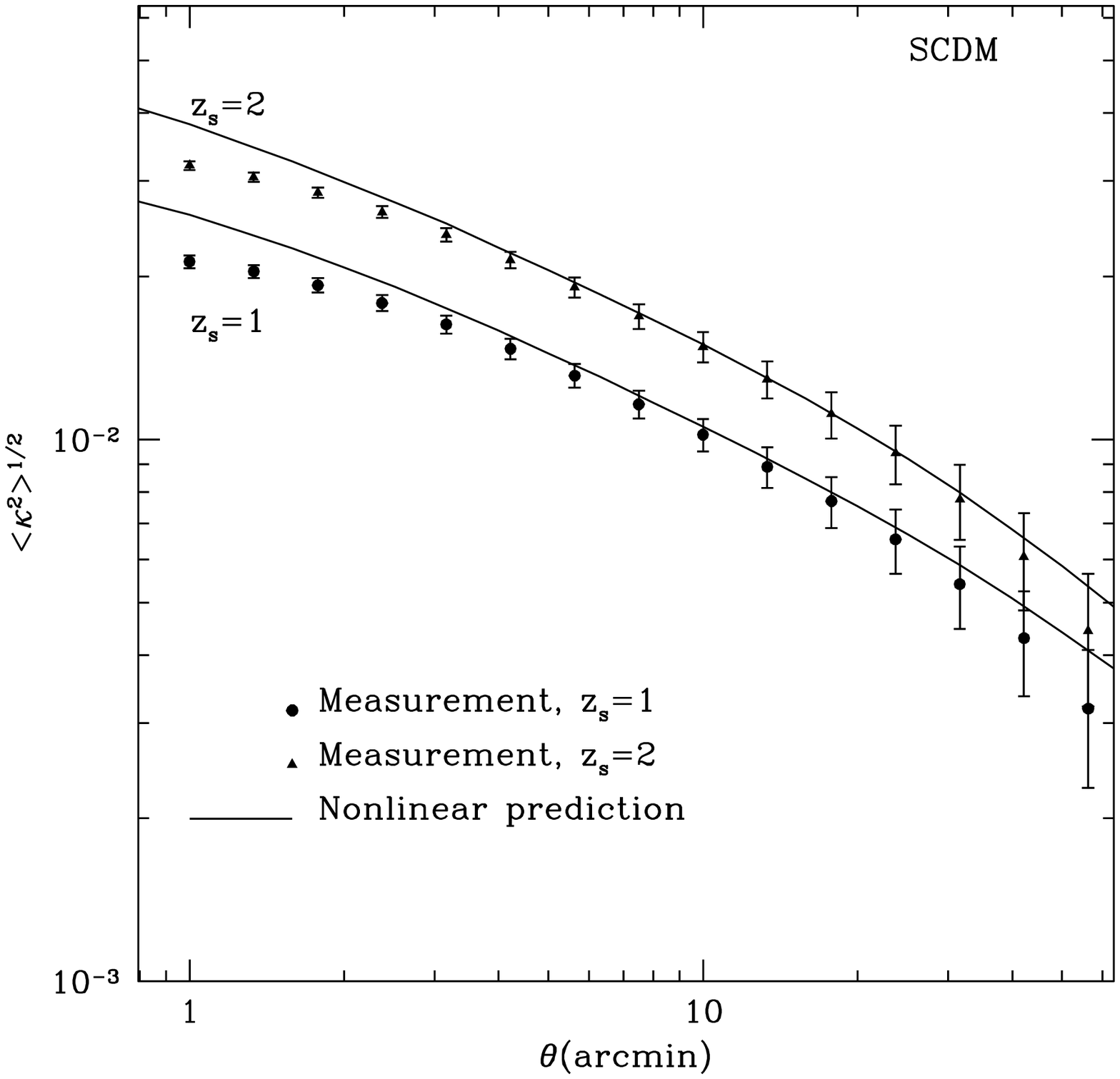,height=7.5cm}
\caption{Root-mean-square of the lensing convergence field filtered by
the top-hat window function as a function of the filter scale.}
\label{fig:rms}
\end{minipage}
\hspace{0.5cm}
\begin{minipage}{7.5cm}
\epsfig{figure=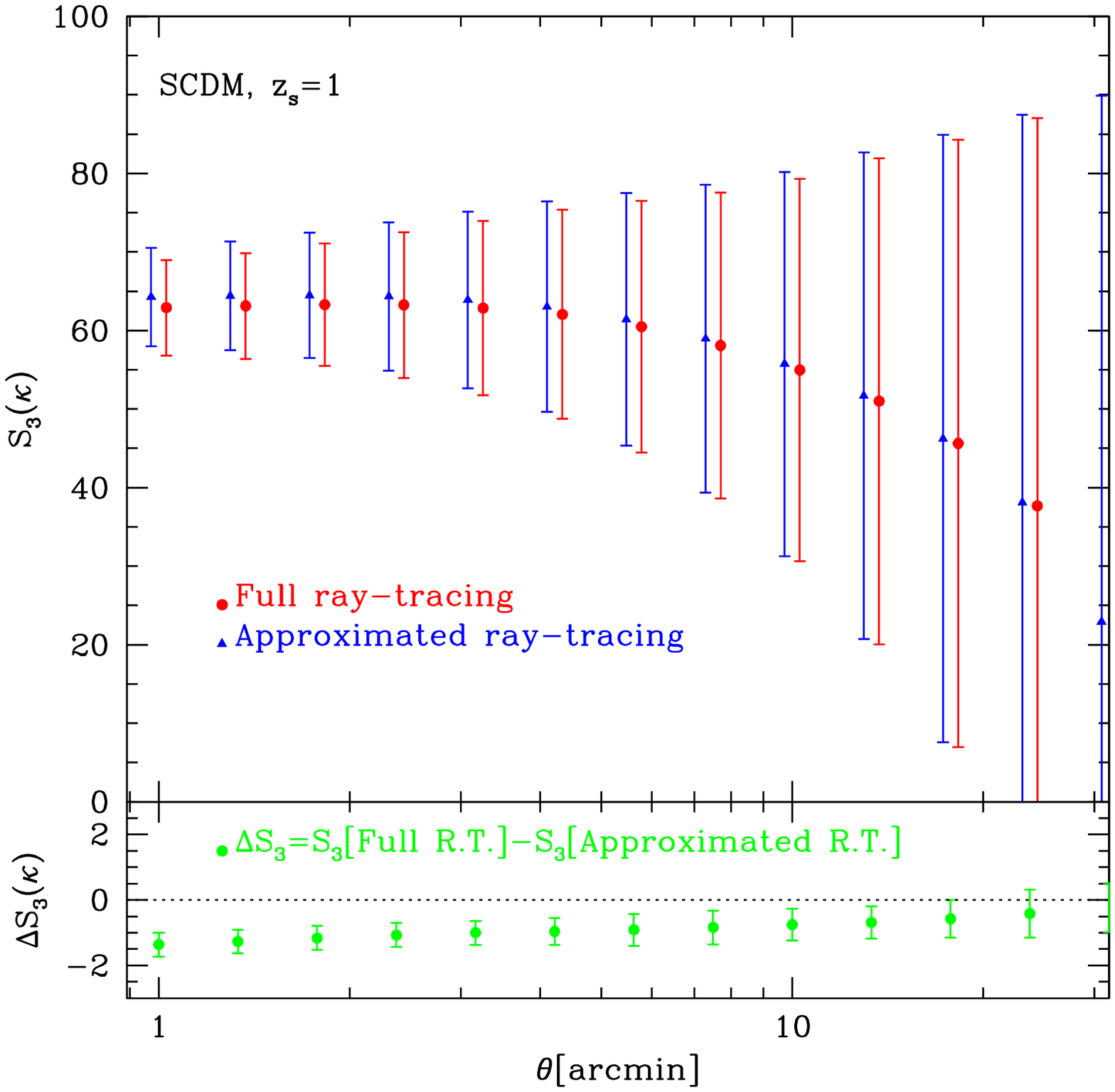,height=7.5cm}
\caption{{\it Top panel}: The skewness parameter $S_3$ as a function
of the filter scale of top-hat window. {\it Bottom panel}: The
difference in $S_3$ between measured from $\kappa$ field of {\it full}
ray-tracing and that of {\it approximated} ray-tracing.}
\label{fig:skew}
\end{minipage}
\end{figure}

Figure \ref{fig:rms} shows the root-mean-square of the top-hat
filtered lensing convergence measured from the ray-tracing simulations
compared with the nonlinear prediction \cite{JA97}.
The dumping on smaller scales comes from the fact that the lack of the
power on scales smaller than the resolution in $N$-body simulation, 
while that on larger scales reflects the finite field
effect, i.e., a lack of powers on scales larger than $N$-body
simulation box.
One may find very good agreement between the measurements and the
nonlinear predictions.

The skewness parameter $S_3$ defined by $S_3 =\langle \kappa^3
\rangle / \langle \kappa^2 \rangle^2$ is known as a powerful probe of
density parameter, $\Omega_m$ \cite{BE97}, 
though, it is difficult to make a precious prediction of its value 
because a description of the nonlinear evolution of the density
bispectrum is required.
The upper panel of Figure \ref{fig:skew} shows $S_3$ measured from the
results of ray-tracing simulations compared with the prediction based
on the quasi-linear theory of density perturbation.
It is clearly shown in Figure 4 that the nonlinearity of the
evolution of density field is very important even at the scale
$\theta\sim10$ arcmin \cite{WA00b}.

In addition to the usual {\it full} ray-tracing simulations,
we performed {\it approximated} ray-tracing simulations to  
examine the effects of the deflections of the light rays
and the lens-lens coupling \cite{BE97} which are neglected in
making the theoretical predictions.
The procedure of the {\it approximated} ray-tracing simulations is
same as that of {\it full} ray-tracing except for artificial
by-passing of both the deflection of light rays and all lens-lens
couplings. 

It is found that the difference in the variance of the lensing
convergence (and of shear) between measured from {\it full} and {\it
approximated} ray-tracing is very small, $|\Delta RMS/RMS| <0.01$.
Therefore the the deflections of the light rays and the lens-lens
coupling can be neglected safely for making the theoretical
predictions of the variance.
For the convergence skewness, we found that the difference is not
significant, $|\Delta S_3 /S_3| <0.05$ for three cosmological models.
These numerical results are compared with the theoretical predictions
and a good agreement is found \cite{WA00b}.

\section*{Acknowledgments}
We would like to thank L. van Waerbeke, F. Bernardeau and A. Thion for
useful discussions.
This research was supported in part by the Direction de la Recherche
du Minist{\`e}re Fran{\c c}ais de la Recherche.  
The computational means (CRAY-98) to do the $N$-body simulations  
were made available to us thanks to the scientific council of 
the Institut du D\'eveloppement et des Ressources en Informatique 
Scientifique (IDRIS)
Numerical computation in this work was partly carried out at the 
the TERAPIX data center.
TH would like to thank Moriond organization for financial supports.


\end{document}